\documentclass[aps,a4paper,amsmath,amssymb,twocolumn,
floatfix,groupedaddress]{revtex4}
\usepackage{bm,amsmath,amssymb,amsfonts}
\usepackage[dvips]{graphicx}
\usepackage{color}
\definecolor{dblue}{rgb}{0.0,0.0,0.7}
\definecolor{dred}{rgb}{0.9,0.0,0.0}
\usepackage{hyperref}

\hypersetup{
    pdfnewwindow=true,      % links in new window
    colorlinks=true,       % false: boxed links; true: colored links
    linkcolor=dred,          % color of internal links
    citecolor=blue,        % color of links to bibliography
    filecolor=blue,      % color of file links
    urlcolor=blue        % color of external links
}
\newcommand{\ie}{{\it i.\,e.}}
\newcommand{\eg}{{\it e.\,g.}}
\newcommand{\etal}{{\it et al.}}
\newcommand{\viz}{{\it viz.}} 
\begin{document}

% Title 
\title{Quasiperiodic magnetic chain as a spin filter for arbitrary spin states}

\author{Biplab Pal}
\email[E-mail:]{ biplabpal@pks.mpg.de, biplabpal2008@gmail.com}
\altaffiliation[Present address: ]
{Department of Physics, Ben-Gurion University, Beer-Sheva 84105, Israel.}
\affiliation{Max Planck Institute for the Physics of Complex Systems, N\"{o}thnitzer Str.\ 38, 
01187 Dresden, Germany }
%
% \date{\today}
%
\begin{abstract}
We show that a quasiperiodic magnetic chain comprising magnetic atomic sites sequenced in Fibonacci pattern 
can act as a prospective candidate for spin filters for particles with arbitrary spin states. This can be 
achieved by tuning a suitable correlation between the amplitude of the substrate magnetic field and the 
on-site potential of the magnetic sites, which can be controlled by an external gate voltage. Such correlation 
leads to a spin filtering effect in the system, allowing one of the spin components to completely pass through 
the system while blocking the others over the allowed range of energies. The underlying mechanism behind this 
phenomena holds true for particles with any arbitrary spin states $S = 1,3/2,2,\hdots$, in addition to the canonical 
case of spin-half particles. Our results open up the interesting possibility of designing a spin demultiplexer 
using a simple quasiperiodic magnetic chain system. Experimental realization of this theoretical study 
might be possible by using ultracold quantum gases, and can be useful in engineering new spintronic devices.       
\end{abstract}

\maketitle

%%%%%%%%%%%%%%%%%%%%%%%%%%%%%%%%%%%%%%%%%%%%%%%%%%%%%%%%%%%%%%%%
\section{Introduction}
The ability to controllably tune, manipulate and detect the spin degree of freedom of a particle 
in low-dimensional systems plays a pivotal role in the field of spintronics~\cite{prinz-science98, 
wolf-sci01, sahoo-np05}. It has emerged as one of the most significant areas of research 
over the past few decades due to its potential to realize new functionalities in future 
electronic devices, and keep the promise to integrate memory and logic in a 
single device. Spin-based electronic devices are assumed to have several important advantages, 
such as high memory storage density, faster access speed, low power consumption, and 
nonvolatility, which give them a significant edge over the existing conventional electronic device 
technologies. To realize such devices for real-life applications, a detailed investigation and understanding 
of the spin-dependent transport in model nanostructured systems is of immense importance, and can 
be treated as a powerful tool to envision the role of the spin degree of freedom in coherent 
electronic systems. Generation of a spin-polarized current source has been one of the key area of 
investigation in the spintronics research domain, and has attracted intense theoretical as well as
experimental research studies over the course of time~\cite{datta-prl02, umansky-sci03, west-prl04, 
umansky-prl03, loss-prl00, kim-apl01, leclair-apl02, brandbyge-prl07, feng-apl10, chuang-nn15, yan-nc16}. 

For the desirable operations and the development of the spin-based devices such as spin-FETs~\cite{datta-apl90}, 
spin-interference devices~\cite{nitta-apl99}, and readout devices for quantum information 
processing and quantum computers~\cite{loss-book02}, the notion of spin-polarized current or 
the so-called spin filter is one of the most pertinent components. Spin interference effects in a quantum ring 
geometry subject to the Rashba spin-orbit interaction was successfully realized experimentally~\cite{nagasawa-prl12} 
a few years ago following an earlier theoretical study~\cite{richter-prb04}. To date, some notable progress 
has been achieved in the study of spin-polarized transport, where people have used ferromagnetic semiconductor 
heterostructures~\cite{fiederling-nat99, ohno-nat99}, metallic multilayer structures~\cite{bauer-prb02}, 
ferromagnetic metal-semiconductor interfaces~\cite{hammar-prl9902}, and carbon-based 
organic materials~\cite{brandbyge-prl07, feng-apl10} among others to achieve highly controllable 
spin-polarized spin injection sources. Furthermore, the study of spin-polarized transport and spin filtering effects in 
quantum networks with loop geometries~\cite{popp-nanotech03, bercioux-prl04, aharony-prb08, pal-scirep16}, or in helical 
molecules~\cite{pan-prb15}, and DNA double helix structure~\cite{guo-prl12, gohler-sci11} has 
also ushered new light into this research arena, revealing different subtleties of spin-polarized 
transport in mesoscopic systems. 

However, to date, the study of spin-polarized transport has mainly focused on the transportation of 
electrons, \ie, for spin $1/2$ particles, while the investigation of spin-polarized transport for particles with higher 
spin states, such as spin $1$ or spin $3/2$ or other higher-order states has not received the same level of attention. 
Only recently, the idea of spin-polarized transport and spin filtering effects for higher spin states in a periodic 
magnetic chain is being proposed and studied in detail~\cite{pal-jpcm16}. We strongly believe that this 
is an area which needs to be explored more rigorously in order to bring out the possibilities of designing 
next-generation novel quantum information storage devices which rely on the spin-polarized transport of particles 
with arbitrary higher spin states. Such systems exhibiting higher-order spin states can be realized in experiments 
using ultracold fermionic or bosonic quantum gases~\cite{ho-prl99, azaria-prl05, pfau-np06, rey-np10, 
takahasi-np12, pagano-np14}. 

It is always an intriguing question to ask whether one can have spin filtering phenomena in a system 
which has no long-range translational order.  In the present article, we address this question and investigate the 
possibility of a spin filtering effect for arbitrary higher-order spin states in a quasiperiodic system. The quasiperiodic 
system we consider for our study is a Fibonacci chain, which represents the simplest model of a quasicrystal~\cite{kohmoto-prl83, 
kohmoto-prb86}. It is well known that, the eigenstates of a periodic system are extended Bloch states~\cite{kittel-book} 
and the corresponding energy spectrum is continuous, while for a disordered system, like in a one-dimensional Anderson 
model, all the eigenstates are localized~\cite{anderson-pr58}. In contrast to these two cases, for a Fibonacci quasiperiodic 
system, the energy spectrum forms a singular continuous Cantor set~\cite{kohmoto-prl83, kohmoto-prb86}. The corresponding 
eigenstates are critical and show multifractal character. Thus quasiperiodic systems, in general, are expected to show up 
poor conducting behavior. In this communication, we present a unique exception of the above scenario and show that, 
for a correlation between the parameters of the Hamiltonian, our simple tight-binding model of a quasiperiodic magnetic chain 
presents a ballistic transmission window for one of the spin channels while completely blocking the particles in the other 
spin channels. It is worth mentioning that, very recently, Mukherjee \etal\ have studied the spin filtering effect in a variety of 
aperiodic systems~\cite{mukherjee-prb18}, where they have taken certain special kinds of quasi-one-dimensional building blocks to 
form the quasiperiodic systems. They have shown that, for some special numerical correlations between the hopping integrals of 
the system, and in some cases an additional external magnetic flux, will lead to a spin filtering effect in the system. We note 
that, in their study, in addition to a major spin channel having a high transmittivity, the other remaining minor spin channels 
also show some transport in their transmission characteristics. In contrast with the above scenario, we propose a very 
simple model of a one-dimensional quasiperiodic magnetic chain. Here one can only tune the values of the on-site potentials 
of the atomic sites by using some external gate voltages to accomplish a complete spin filtering effect for one of the desired spin 
components (channels), while the remaining spin components will have zero transport through the system under this condition.    

In what follows, we present the model and describe the essential results. The layout of the paper is the following. 
In Sec.~\ref{model}, we introduce our model and describe the essential mathematical framework employed to 
extract the results. The results for the spin-dependent transport and the spin filtering effect along with the corresponding 
local density of states (LDOS) for different spin channels are discussed in detail in Sec.~\ref{results}. Finally, in 
Sec.~\ref{conclu}, we draw our conclusion with a summary of the key findings and their possible applications.
%######################################################
\begin{figure}[ht]
\includegraphics[clip,width=\columnwidth]{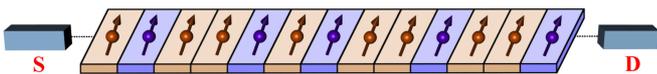}
\caption{Schematic diagram of a finite-size Fibonacci sequenced quasiperiodic 
magnetic layered structure coupled between two semi-infinite nonmagnetic 
leads, \viz, source (S) and drain (D).}
\label{fig:QP-mag-chain}
\end{figure}
%######################################################          
%%%%%%%%%%%%%%%%%%%%%%%%%%%%%%%%%%%%%%%%%%%%%%%%%%%%%%%%%%%%%%%%
\section{The model and the theoretical framework}
\label{model}
\subsection{The Fibonacci chain}
We propose the construction of a linear magnetic chain model following a Fibonacci sequence. The Fibonacci sequence is 
a quasiperiodic sequence of two letters; say, $\mathcal{A}$ and $\mathcal{B}$. To construct the system, one can start 
from the letter $\mathcal{A}$, and then apply the following substitution rule to grow the system in to its different 
higher-order generations:
%---------------------------------------------------------------------  
\begin{equation}
\mathcal{A} \rightarrow \mathcal{A}\mathcal{B} \ \text{and} \ \mathcal{B} \rightarrow \mathcal{A}.
\label{eq:fiborule}
\end{equation}
%---------------------------------------------------------------------  
By using the above prescription in Eq.~\eqref{eq:fiborule}, we can easily construct the different generations of a Fibonacci 
system as follows, $\mathcal{A}\mathcal{B}\mathcal{A}$, $\mathcal{A}\mathcal{B}\mathcal{A}\mathcal{A}\mathcal{B}$, 
$\mathcal{A}\mathcal{B}\mathcal{A}\mathcal{A}\mathcal{B}\mathcal{A}\mathcal{B}\mathcal{A}$, and so on. For our 
model, the letters $\mathcal{A}$ and $\mathcal{B}$ are simply replaced by two kinds of magnetic atomic sites grafted on 
some substrates to form the system shown in Fig.~\ref{fig:QP-mag-chain}. These magnetic sites have magnetic 
moments $\vec{h}_{\mathcal{A}}$ and $\vec{h}_{\mathcal{B}}$ respectively, associated with them. In the thermodynamic 
limit, \ie, for an infinitely long chain, the ratio of $\mathcal{A}$-type to $\mathcal{B}$-type magnetic sites is 
incommensurate and gives $\sigma = (1+\sqrt{5})/2$, which is known as the `golden ratio'. 
\subsection{Hamiltonian of the system}
\begin{figure}[ht]
\includegraphics[clip,width=0.5\columnwidth]{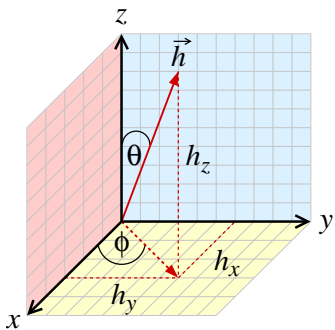}
\caption{Decomposition of $\vec{h}$ in a three-dimensional plane. $\theta$ denotes the 
polar angle and $\phi$ denotes the azimuthal angle.}
\label{fig:h-vec}
\end{figure}

The Hamiltonian of the system in a tight-binding framework can be written as, 
%---------------------------------------------------------------------  
\begin{equation}
\bm{H}=\sum_{i} \bm{c}_{i}^{\dagger} \left(\bm{\epsilon}_{i} - 
\vec{h}_{i} \cdot \vec{\bm{\mathcal{S}}}^{(S)}_{i} \right) \bm{c}_{i} + 
\sum_{\langle i,j \rangle} \Big(\bm{c}_{i}^{\dagger} \bm{t}_{ij} 
\bm{c}_{j} + \text{H.c.}\Big), 
\label{eq:hamiltonian}
\end{equation}
%---------------------------------------------------------------------  
where $\langle i,j \rangle$ indicates the nearest-neighbor atomic sites. We note that, each of the terms 
$\bm{c}_{i}^{\dagger}(\bm{c}_{i})$, $\bm{\epsilon}_{i}$, $\bm{t}_{ij}$, and $\vec{\bm{\mathcal{S}}}^{(S)}_{i}$ 
represents multi-component matrices with dimensions that depend on the spin of the particles. For the simplest 
case of a spin-half ($S=1/2$) particle, these matrices, \viz, creation (annihilation) matrix, on-site potential matrix, 
and hopping matrix, take the following forms:
%---------------------------------------------------------------------  
\begin{eqnarray}
&\bm{c}_{i}^{\dagger} =   
\left ( \begin{array}{cc}
c^{\dagger}_{i,\uparrow} & c^{\dagger}_{i,\downarrow}  
\end{array} \right ),\
\bm{c}_{i} =   
\left ( \begin{array}{c}
c_{i,\uparrow} \\ c_{i,\downarrow}  
\end{array} \right ),\
\bm{\epsilon}_{i} = 
\left( \begin{array}{cccc}
\epsilon_{i,\uparrow} & 0 \\ 
0 & \epsilon_{i,\downarrow} 
\end{array}\right), \nonumber \\
&\text{and}\ 
\bm{t}_{ij} = 
\left( \begin{array}{cccc}
t & 0 \\ 
0 & t
\end{array}\right), 
\end{eqnarray}
%---------------------------------------------------------------------    
where the indices `$\uparrow$' and `$\downarrow$' refer to the spin-{\it up} and spin-{\it down} 
components (`channels'), respectively. Note that the dimension of these matrices will increase 
proportionately as we go to the higher-order spin cases, \viz, $S=1,3/2,\hdots$, and so on. The term 
$\vec{h}_{i} \cdot \vec{\bm{\mathcal{S}}}^{(S)}_{i}$ represents the interaction of the spin ($S$) of the 
injected particle with the local magnetic field $\vec{h}_{i}\equiv \left(h_{x},h_{y},h_{z}\right)$ 
at site $i$. $\vec{\bm{\mathcal{S}}}^{(S)}_{i}$ represents the set of {\it generalized} Pauli spin matrices 
$\left(\bm{\mathcal{S}}_{x}, \bm{\mathcal{S}}_{y}, \bm{\mathcal{S}}_{z}\right)$ expressed in units of 
$\hbar S$ for an incoming particle with spin $S$. For the spin-half ($S=1/2$) case, 
$\left(\bm{\mathcal{S}}_{x}, \bm{\mathcal{S}}_{y}, \bm{\mathcal{S}}_{z}\right)$ turns out to be the set of 
the usual Pauli spin matrices $\left(\bm{\sigma}_{x},\bm{\sigma}_{y},\bm{\sigma}_{z}\right)$, and the term 
$\vec{h}_{i} \cdot \vec{\bm{\mathcal{S}}}^{(S)}_{i}$ at site $i$, will have the following explicit form: 
%---------------------------------------------------------------------
\begin{equation}
\vec{h}_{i} \cdot \vec{\bm{\mathcal{S}}}^{(S=1/2)}_{i}
= \left(\def\arraystretch{1.5}\begin{array}{cc}
h_{i} \cos\theta_{i}   &   h_{i} \sin\theta_{i} e^{-i\phi_{i}} \\
h_{i} \sin\theta_{i} e^{i\phi_{i}}   &   -h_{i} \cos\theta_{i} 
\end{array} \right),
\label{eq:spinflip}
\end{equation}
%---------------------------------------------------------------------
where $h_{i}$ is the amplitude of the vector $\vec{h}_{i}$, and $\theta_{i}$ and $\phi_{i}$ denote the polar and 
azimuthal angles, respectively, as shown in Fig.~\ref{fig:h-vec}. 
%######################################################
%###################################################### 
\subsection{Equivalence of the spin system with a multi-strand ladder network}
Using the Hamiltonian in Eq.~\eqref{eq:hamiltonian}, one can write down the time-independent Schr\"{o}dinger 
equation for a general spin $S$ system as follows:
%---------------------------------------------------------------------
\begin{equation}
H |\Psi\rangle_{S} = E |\Psi\rangle_{S}, \quad \text{with}\ 
|\Psi\rangle_{S}=\sum_{\ell}\sum_{m_{S}=-S}^{+S}\psi_{\ell,m_{S}}|\ell,m_{S}\rangle. 
\label{eq:Sch-eqn}
\end{equation}
%---------------------------------------------------------------------
A simplification of Eq.~\eqref{eq:Sch-eqn} for a spin-half ($S=1/2$) system will lead to the following set of difference 
equations corresponding to the spin-{\it up} ($\uparrow$) and spin-{\it down} ($\downarrow$) channels, respectively, 
as follows:
%--------------------------------------------------------------------- 
\begin{subequations}
\label{eq:spinhalf-diff}
\begin{multline}
\big[E- \big( \epsilon_{i,\uparrow} - h_{i}\cos \theta_{i} \big) \big] \psi_{i,\uparrow} + 
h_{i} \sin\theta_{i} e^{-i\phi_{i}} \psi_{i,\downarrow} \\
= t \psi_{i+1,\uparrow} + t \psi_{i-1,\uparrow},
\label{eq:spinhalf-diff-up}
\end{multline}
%%%%%%%%%%%%%%%%%%%%%%%%%%%%
\begin{multline}
\big[E- \big( \epsilon_{i,\downarrow} + h_{i}\cos\theta_{i} \big) \big] \psi_{i,\downarrow} +
h_{i} \sin\theta_{i} e^{i\phi_{i}} \psi_{i,\uparrow} \\
= t \psi_{i+1,\downarrow} + t \psi_{i-1,\downarrow}.
\label{eq:spinhalf-diff-down}
\end{multline}
\end{subequations}
%---------------------------------------------------------------------
It is interesting to note that, the above set of difference equations Eq.~\eqref{eq:spinhalf-diff-up} and~\eqref{eq:spinhalf-diff-down} 
resemble the difference equations for a spinless particle in a two-strand ladder network~\cite{sil-prb08}. The effective on-site 
potentials for the {\it upper} strand (identified with the spin-{\it up} ($\uparrow$) component) and the {\it lower} strand 
(identified with the spin-{\it down} ($\downarrow$) component) of the analogous ladder network are 
$\epsilon_{i,\uparrow} - h_{i}\cos \theta_{i}$ and $\epsilon_{i,\downarrow} + h_{i}\cos\theta_{i}$ respectively, the hopping 
amplitude between the two neighboring sites along each strand of the ladder can be identified as $t$, while the term 
$h_{i} \sin\theta_{i} e^{i\phi_{i}}$ plays the role of the interstrand coupling along the $i$-th rung of the ladder. 

In a similar way, starting from the Eq.~\eqref{eq:Sch-eqn} for a spin-$1$ ($S=1$) system, we can obtain the following set of 
three coupled difference equations for the three spin channels $1$, $0$, and $-1$ respectively, as 
%--------------------------------------------------------------------- 
\begin{subequations}
\label{eq:spin1-diff}
\begin{multline}
\big[E - \big( \epsilon_{i,1} - h_{i}\cos \theta_{i} \big) \big] \psi_{i,1} + 
\dfrac{1}{\sqrt{2}} h_{i}\sin \theta_{i}e^{-i\phi_{i}} \psi_{i,0} \\
= t \psi_{i+1,1} + t \psi_{i-1,1},
\label{eq:spin1-diff-1}
\end{multline}
%%%%%%%%%%%%%%%%%%%%%%%%%%%%
\begin{multline}
\big[E - \epsilon_{i,0} \big] \psi_{i,0} +  
\dfrac{1}{\sqrt{2}} h_{i}\sin \theta_{i}e^{i\phi_{i}} \psi_{i,1} + 
\dfrac{1}{\sqrt{2}} h_{i}\sin \theta_{i}e^{-i\phi_{i}} \psi_{i,-1} \\
=  t \psi_{i+1,0} + t \psi_{i-1,0},
\label{eq:spin1-diff-0}
\end{multline}
%%%%%%%%%%%%%%%%%%%%%%%%%%%%
\begin{multline}
\big[E - \big( \epsilon_{i,-1} + h_{i}\cos\theta_{i} \big) \big] \psi_{i,-1} + 
\dfrac{1}{\sqrt{2}} h_{i}\sin \theta_{i}e^{i\phi_{i}} \psi_{i,0} \\
= t \psi_{i+1,-1} + t \psi_{i-1,-1},
\label{eq:spin1-diff-m1}
\end{multline}
\end{subequations}
%--------------------------------------------------------------------- 
where we have used the following set of spin matrices 
$\left(\bm{\mathcal{S}}^{S=1}_{x}, \bm{\mathcal{S}}^{S=1}_{y}, \bm{\mathcal{S}}^{S=1}_{z}\right)$ 
for a spin-$1$ system:
\begin{eqnarray}
& \bm{\mathcal{S}}^{S=1}_{x}
= \dfrac{1}{\sqrt{2}} \left(\def\arraystretch{1.5}\begin{array}{ccc}
0  &  1  &  0 \\
1  &  0  &  1 \\
0  &  1  &  0 
\end{array} \right), 
\bm{\mathcal{S}}^{S=1}_{y}
= \dfrac{1}{\sqrt{2}} \left(\def\arraystretch{1.5}\begin{array}{ccc}
0  &  -i  &  0 \\
i  &  0  &  -i \\
0  &  i  &  0 
\end{array} \right), \nonumber \\
& \text{and}\ \bm{\mathcal{S}}^{S=1}_{z}
= \left(\def\arraystretch{1.5}\begin{array}{ccc}
1  &  0  &  0 \\
0  &  0  &  0 \\
0  &  0  &  -1 
\end{array} \right).
\label{eq:spin1-matrices}
\end{eqnarray} 
Once again, the above prescription leads to the fact that, a spin-$1$ ($S=1$) system can be identified with a 
three-strand ladder network for a spinless particle. The above two analyses for the spin-half ($S=1/2$) and the spin-1 ($S=1$) 
cases lead to the conclusion that, the above treatment can be extended to a general spin-$S$ system and can be identified 
with an equivalent ($2S+1$) strand ladder model for spinless particles. We note that, as we go to the higher-order spin cases, 
there will be more such coupled difference equations corresponding to the different spin channels. This analogy between 
the spin model and the multi-strand ladder network is employed to engineer the spin filtering effect for our quasiperiodic 
system, as described in the subsequent sections.
%%%%%%%%%%%%%%%%%%%%%%%%%%%%%%%%%%%%%%%%%%%%%%%%%%%%%%%%%%%%%%%%
\section{Results and discussion}
\label{results}
One of the essential requirements to have the spin filtering effect, is to decouple the different spin channels from each other, 
\ie, there should not be any spin mixing between different spin components. For our model, this condition can be satisfied 
by setting the polar angle $\theta_{i} = 0$ $\forall$ $i$. By looking at the set of equations~\eqref{eq:spinhalf-diff} 
and~\eqref{eq:spin1-diff}, it can be easily understood that the hybridization terms (spin-mixing terms) between different 
spin components vanishes for $\theta_{i} = 0$ as the $\sin\theta_{i}$ terms vanishes under this condition irrespective of the 
value of the azimuthal angle $\phi_{i}$. The physical meaning of the above condition is that, the magnetic moments of the 
atomic sites in the system have to be aligned along the $z$ axis parallel to each other.
\subsection{Spin-half ($S=1/2$) system} 
For a spin-half ($S=1/2$) system, the above choice of $\theta_{i}$ will lead to the following set of equations from 
Eq.~\eqref{eq:spinhalf-diff}:
%--------------------------------------------------------------------- 
\begin{subequations}
\label{eq:spinhalf-decouple}
\begin{multline}
\big[E- \big( \epsilon_{i,\uparrow} - h_{i}\big) \big] \psi_{i,\uparrow}  
= t \psi_{i+1,\uparrow} + t \psi_{i-1,\uparrow},
\label{eq:spinhalf-decouple-up}
\end{multline}
%%%%%%%%%%%%%%%%%%%%%%%%%%%%
\begin{multline}
\big[E- \big( \epsilon_{i,\downarrow} + h_{i}\big) \big] \psi_{i,\downarrow} 
= t \psi_{i+1,\downarrow} + t \psi_{i-1,\downarrow}.
\label{eq:spinhalf-decouple-down}
\end{multline}
\end{subequations}
%---------------------------------------------------------------------
It is apparent from the two equations above that the spin-{\it up} ($\uparrow$) and the spin-{\it down} ($\downarrow$) 
channels are now completely decoupled from each other. We furthermore can choose 
$\epsilon_{i,\uparrow}=\epsilon_{i,\downarrow}=\epsilon_{i}$, \ie, the on-site energies at an $i$-th atomic site are to
be the same for both the spin-up ($\uparrow$) and the spin-down ($\downarrow$) particles. For our model, $h_{i}$ can 
take two possible values $h_{\mathcal{A}}$ and $h_{\mathcal{B}}$ sequenced following a Fibonacci pattern as depicted in 
Fig.~\ref{fig:QP-mag-chain}. Such a sequence can be generated mathematically by using
%---------------------------------------------------------------------
\begin{equation}
h_{i} = P + Q \big( \lfloor (i+1) (\sigma-1) \rfloor - \lfloor i (\sigma-1) \rfloor \big),
\label{eq:fibo-formula}
\end{equation} 
where the function $\lfloor x \rfloor$ denotes the greatest integer lower than $x$, $\sigma = (1+\sqrt{5})/2$, and 
$P$ and $Q$ are two parameters that control the values of $h_{\mathcal{A}}$ and $h_{\mathcal{B}}$. 
Using Eq.~\eqref{eq:fibo-formula}, one can easily find out that, the values of $h_{\mathcal{A}}$ and $h_{\mathcal{B}}$ 
will turn out to be $h_{\mathcal{A}}=P+Q$ and $h_{\mathcal{B}}=P$ respectively, sequenced in a Fibonacci pattern. 
We can also choose the values of the on-site potentials $\epsilon_{i}$ to follow a Fibonacci pattern, consisting of two kinds 
of constituents $\epsilon_{\mathcal{A}}$ and $\epsilon_{\mathcal{B}}$ respectively. The values of these two on-site 
potentials can be easily controlled by using an external gate voltage~\cite{umansky-sci03}. Hence, one can easily 
have the exactly identical Fibonacci pattern for $\epsilon_{i}$ as that of $h_{i}$. 

Now with this convention, if we set 
$\epsilon_{\mathcal{A}}=\Delta+h_{\mathcal{A}}$ and $\epsilon_{\mathcal{B}}=\Delta+h_{\mathcal{B}}$ 
(where $\Delta$ is some constant value which sets the center of the energy spectrum), 
then from Eq.~\eqref{eq:spinhalf-decouple}, it immediately follows that, for the spin-{\it up} 
($\uparrow$) channel, the effective on-site potentials on different atomic sites will have a constant value, while for 
the spin-{\it down} ($\downarrow$) channel the effective on-site potentials on different atomic sites will follow a 
Fibonacci quasiperiodic pattern. As a result of this, we will have an absolutely continuous energy spectrum for 
the spin-{\it up} ($\uparrow$) channel populated with extended states while the spin-{\it down} ($\downarrow$) 
channel will feature a singular continuous multifractal spectrum. Thus, under this condition, we will have a high 
transmission probability for the spin-{\it up} ($\uparrow$) particles whereas the spin-{\it down} ($\downarrow$) 
particles will encounter zero transmission probability. To analyze this fact, we evaluate the local density of 
states (LDOS) for the different spin channels by using the Green's function technique. The formula for the LDOS is,
\begin{equation}
\rho_{j, m_{S}} = -\dfrac{1}{\pi}\lim_{\eta \rightarrow 0^+} 
\big[ \textrm{Im}\big(\langle j,m_{S}|\bm{G}(E)| 
j,m_{S}\rangle\big) \big],
\label{eq:Green}
\end{equation} 
where $\bm{G}(E)=(z^+ \bm{I} - \bm{H})^{-1}$ is the Green's function with $z^+=E+i \eta$ ($\eta \rightarrow 0^+$), 
and $m_{S}$ will have $2S+1$ values for a general spin $S$ system, \eg, for the spin-half ($S=1/2$) case, 
$m_{S}=1/2(\uparrow),-1/2(\downarrow)$. 
%######################################################
\begin{figure}[ht]
\includegraphics[clip,width=0.49\columnwidth]{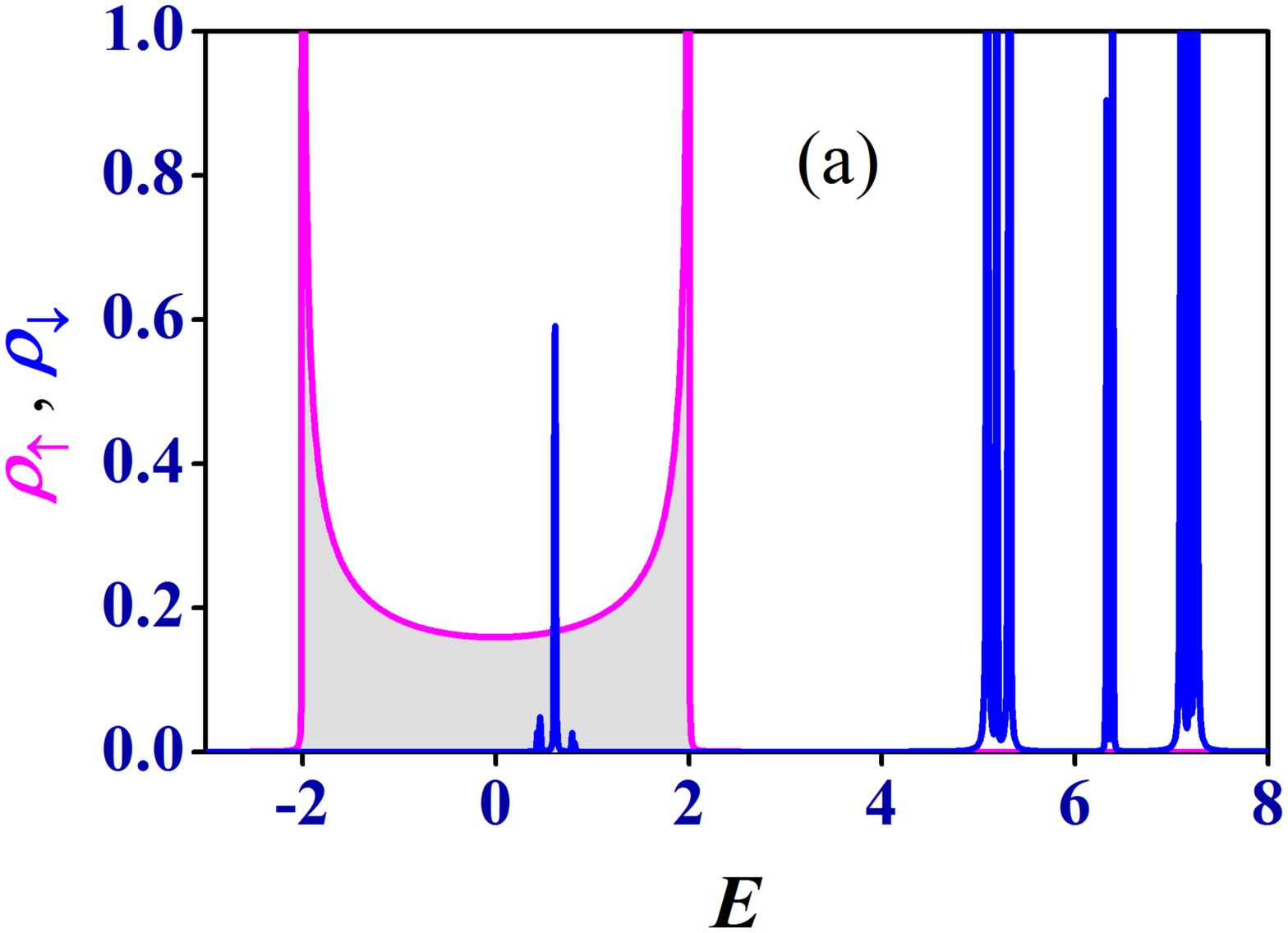}
\includegraphics[clip,width=0.49\columnwidth]{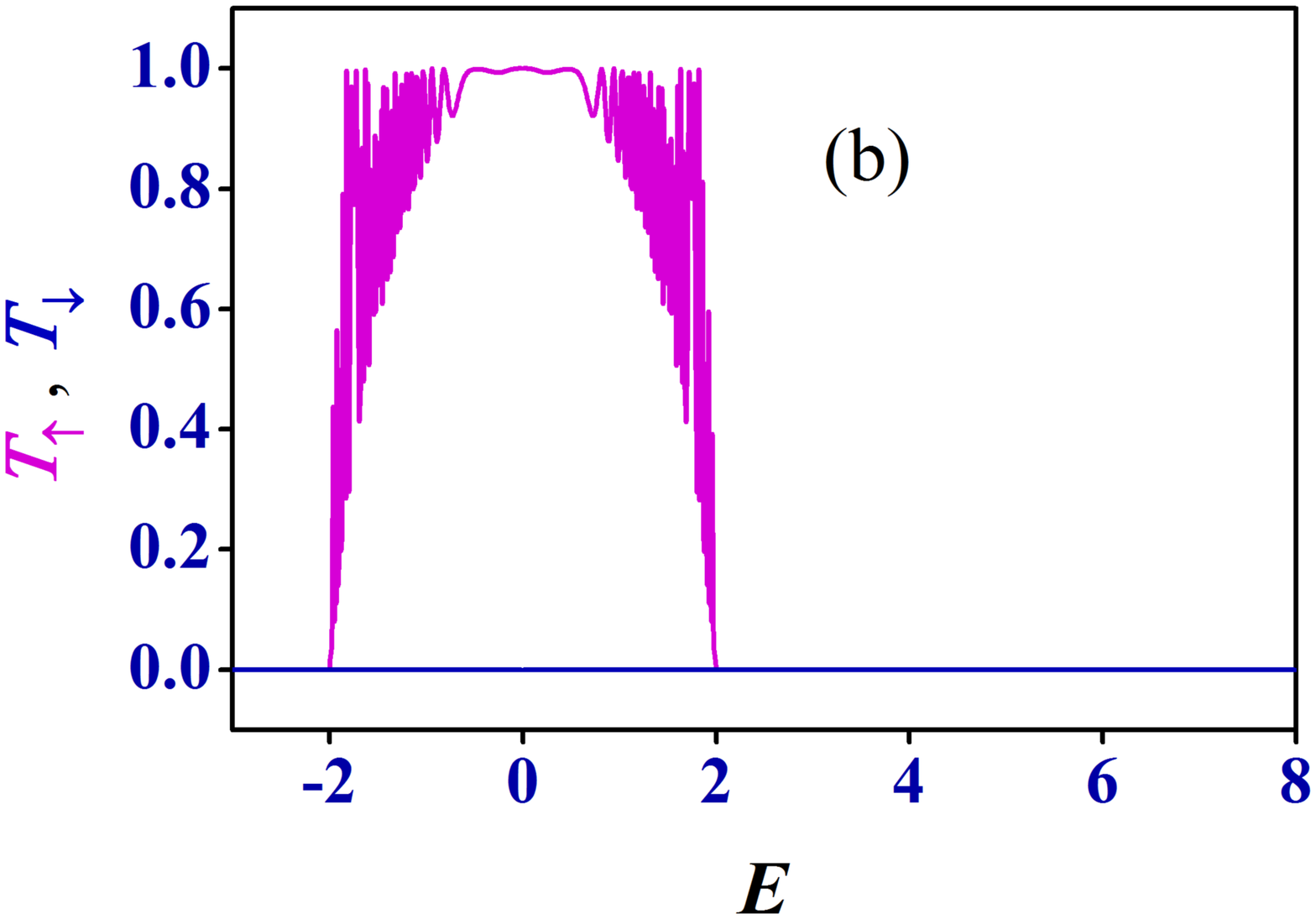}
\caption{(a) Plots of LDOS for the spin-{\it up} ($\uparrow$) and the 
spin-{\it down} ($\downarrow$) channels under the correlation condition 
$\epsilon_{\mathcal{A}}=\Delta + h_{\mathcal{A}}$ and 
$\epsilon_{\mathcal{B}}=\Delta + h_{\mathcal{B}}$. The spin-{\it up} 
($\uparrow$) channel exhibits an absolutely continuous spectrum 
(shaded portion) while the spin-{\it down} ($\downarrow$) channel 
shows a multifractal singular continuous spectrum. We set $\Delta=0$, 
$h_{\mathcal{A}}=3$, and $h_{\mathcal{B}}=0.5$ 
measured in units of the hopping integral $t$. (b) The corresponding 
transmission probabilities $T_{\uparrow}$ and $T_{\downarrow}$ 
for the spin-{\it up} ($\uparrow$) and the 
spin-{\it down} ($\downarrow$) components computed for a $15$-th 
generation Fibonacci magnetic chain with $610$ atomic sites.}
\label{fig:ldos-trans-spinhalf-1}
\end{figure}
%######################################################  

We show the plots of LDOS for the spin-{\it up} ($\uparrow$) and the spin-{\it down} ($\downarrow$) channels in 
Fig.~\ref{fig:ldos-trans-spinhalf-1}(a). 
We have used a real-space renormalization group (RSRG) method~\cite{pal-physE14, pal-pla14} to compute the LDOS 
spectrum for the different spin channels. We can clearly observe that, the spin-{\it up} ($\uparrow$) channel shows 
an absolutely continuous energy spectrum in between $E=\Delta-2t$ and $E=\Delta+2t$ (here we have taken $\Delta=0$ and 
$t=1$). All the eigenstates populated under this absolutely continuous energy spectrum are of extended character. For the 
spin-{\it down} ($\downarrow$) channel, we have a multifractal energy spectrum with self-similarity, which exhibits the 
signature of a quasiperiodic system. The corresponding transmission probabilities for the two spin channels, \viz, up ($\uparrow$) 
and down ($\downarrow$), are exhibited in Fig.~\ref{fig:ldos-trans-spinhalf-1}(b). Evidently, we have a high transmission 
probability ($T_{\uparrow}$) for the spin-{\it up} ($\uparrow$) component corresponding to the absolutely continuous 
spectrum in Fig.~\ref{fig:ldos-trans-spinhalf-1}(a), while the spin-{\it down} ($\downarrow$) component gets completely blocked with 
zero transmission probability ($T_{\downarrow}$). To evaluate these transmission characteristics, we take a finite-size 
quasiperiodic magnetic chain and couple it in between two nonmagnetic periodic leads, \viz, source (S) and drain (D), as 
shown schematically in Fig.~\ref{fig:QP-mag-chain}. For the results of the transmission probabilities presented here, we 
have considered a $15$-th generation Fibonacci chain with $610$ atomic sites. The values of the other parameters; 
namely, the on-site potentials for the atomic sites in the leads, the hopping amplitudes for sites in the lead, and the lead to 
magnetic chain (MC) couplings are chosen to be $\epsilon_{S}=\epsilon_{D}=0$, $t_{S}=t_{D}=4$, and 
$t_{S,MC}=t_{MC,D}=4$, respectively, for our calculation.  We have used a standard transfer-matrix method (TMM) 
elaborated in detail in Ref.~\cite{pal-jpcm16} to obtain the transmission probabilities corresponding to the different 
spin components for our quasiperiodic system.
%######################################################
\begin{figure}[ht]
\includegraphics[clip,width=0.49\columnwidth]{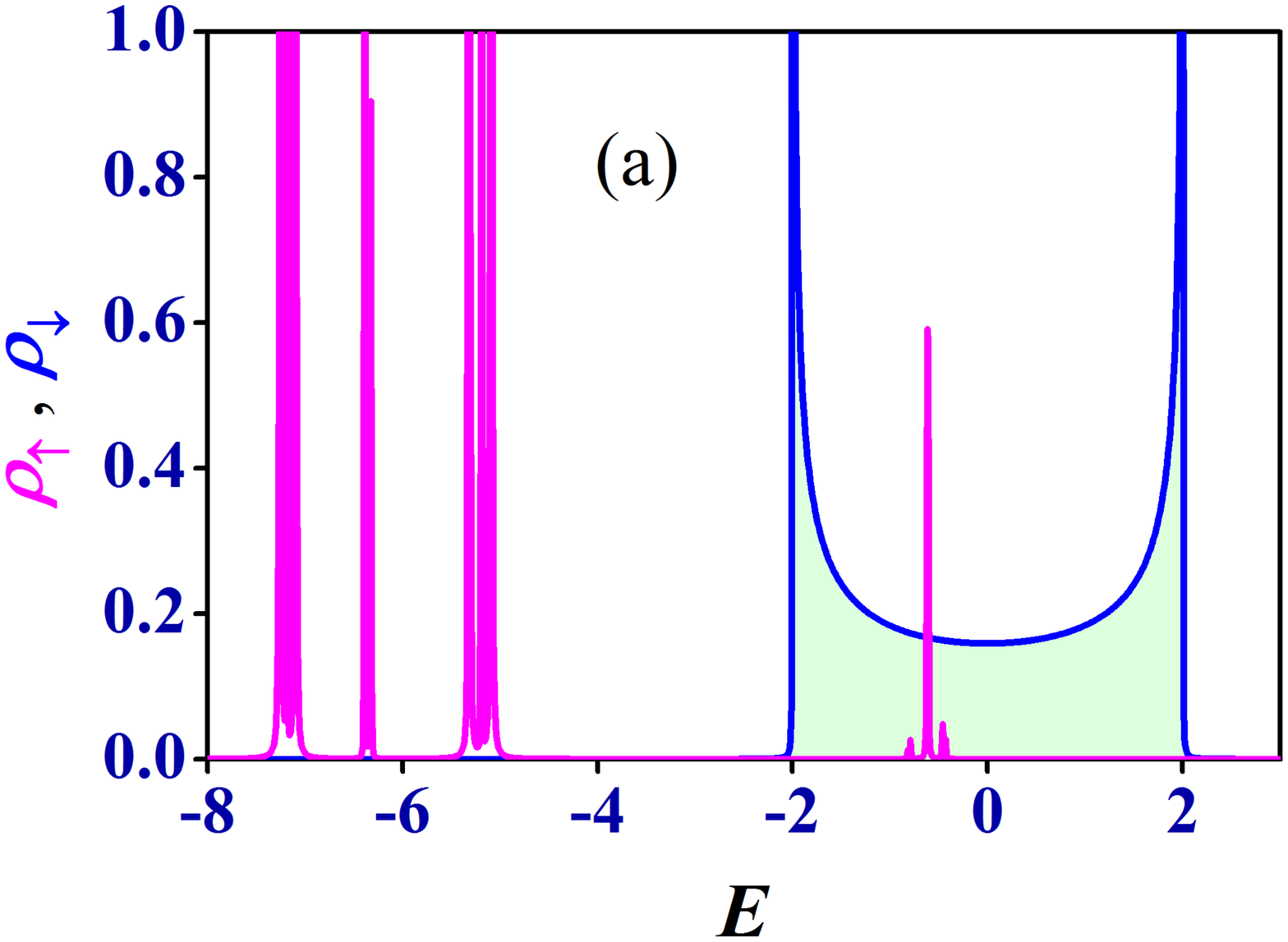}
\includegraphics[clip,width=0.49\columnwidth]{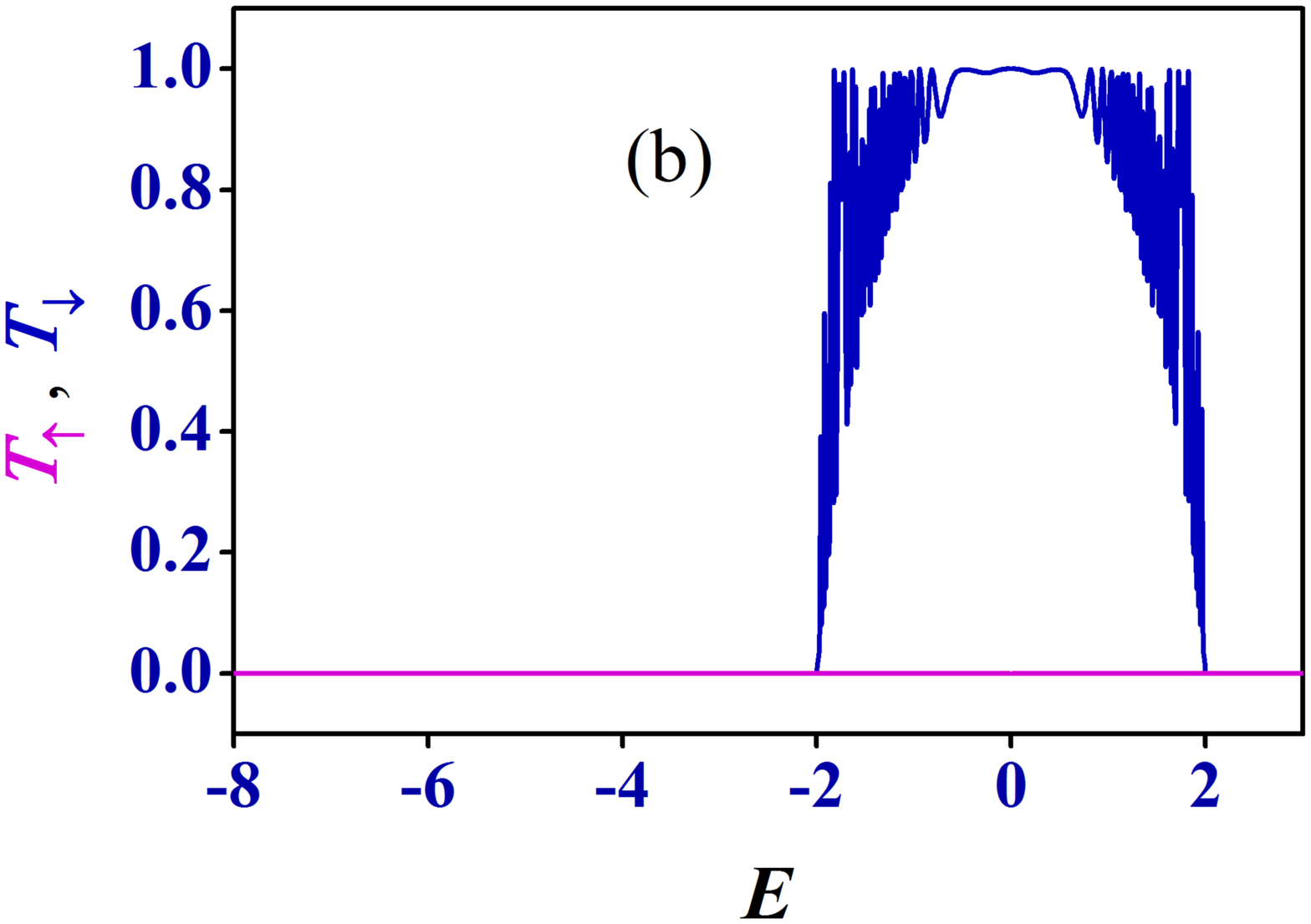}
\caption{(a) Plots of LDOS for the spin-{\it up} ($\uparrow$) and the 
spin-{\it down} ($\downarrow$) components with the correlation condition 
$\epsilon_{\mathcal{A}}=\Delta - h_{\mathcal{A}}$ and 
$\epsilon_{\mathcal{B}}=\Delta - h_{\mathcal{B}}$. We choose $\Delta=0$, 
$h_{\mathcal{A}}=3$, and $h_{\mathcal{B}}=0.5$ 
measured in units of the hopping integral $t$. (b) The corresponding 
transmission characteristics $T_{\uparrow}$ and $T_{\downarrow}$ 
measured for a $15$-th generation Fibonacci magnetic chain containing 
$610$ atomic sites. The other parameters are same as 
in Fig.~\ref{fig:ldos-trans-spinhalf-1}.}
\label{fig:ldos-trans-spinhalf-2}
\end{figure}
%###################################################### 

We note that, one can have exactly the opposite phenomenon as compared with the results described in the 
last two paragraphs for a different choice of correlation between the two sets of parameters 
$\{ \epsilon_{\mathcal{A}}, \epsilon_{\mathcal{B}} \}$ and $\{ h_{\mathcal{A}}, h_{\mathcal{B}} \}$, 
\viz, $\epsilon_{\mathcal{A}}=\Delta-h_{\mathcal{A}}$ and $\epsilon_{\mathcal{B}}=\Delta-h_{\mathcal{B}}$, 
where $\Delta$ is a constant value which sets the center of the energy spectrum. With this choice of the correlation, 
it follows from Eq.~\eqref{eq:spinhalf-decouple} that, now we will have a constant value of the effective on-site 
energies for the spin-{\it down} ($\downarrow$) channel whereas the particles in the spin-{\it up} ($\uparrow$) 
channel will feel a quasiperiodic effective on-site potential. Consequently, we will have an absolutely continuous 
energy spectrum for the spin-{\it down} ($\downarrow$) channel and a multifractal self-similar singular continuous 
spectrum for the spin-{\it up} ($\uparrow$) channel as shown in Fig.~\ref{fig:ldos-trans-spinhalf-2}(a). 
The resulting transmission characteristics for this case are displayed in Fig.~\ref{fig:ldos-trans-spinhalf-2}(b), 
where we can clearly see that, the particles with spin-{\it down} ($\downarrow$) component will have a 
transparent transmitting window for the allowed energy regime while the particles with spin-{\it up} 
($\uparrow$) component will have a completely opaque transmitting window. So the conclusion is that, one can 
make a tunable spin filter for one of the desired spin components by choosing an appropriate correlation between 
$\{ \epsilon_{\mathcal{A}}, \epsilon_{\mathcal{B}} \}$ and $\{ h_{\mathcal{A}}, h_{\mathcal{B}} \}$. This can 
be achieved basically by suitably tuning the values of $ \epsilon_{\mathcal{A}}$ and $ \epsilon_{\mathcal{B}}$ 
using some external gate voltages. The typical experimental value of the spacial extension over 
which a modulation of the gate voltage can be achieved is in the range of 100--150 nm.        
\subsection{Spin-1 ($S=1$) system} 
Now we turn to the case of a spin-1 ($S=1$) system which has three components, \viz, $1$, $0$, and $-1$. It can be 
identified with a $2S+1=3$ strand ladder network. Once again, by setting the polar angle $\theta_{i} = 0$ $\forall$ $i$, 
we can decouple the three spin channels from each other, and analyze the transport properties for each of these three 
different spin channels. With the above choice, the three decoupled equations following from Eq.~\eqref{eq:spin1-diff} 
can be written as,
%--------------------------------------------------------------------- 
\begin{subequations}
\label{eq:spin1-decouple}
\begin{multline}
\big[E - \big( \epsilon_{i,1} - h_{i} \big) \big] \psi_{i,1} 
= t \psi_{i+1,1} + t \psi_{i-1,1},
\label{eq:spin1-decouple-1}
\end{multline}
%%%%%%%%%%%%%%%%%%%%%%%%%%%%
\begin{multline}
\big[E - \epsilon_{i,0} \big] \psi_{i,0} 
=  t \psi_{i+1,0} + t \psi_{i-1,0},
\label{eq:spin1-decouple-0}
\end{multline}
%%%%%%%%%%%%%%%%%%%%%%%%%%%%
\begin{multline}
\big[E - \big( \epsilon_{i,-1} + h_{i} \big) \big] \psi_{i,-1} + 
= t \psi_{i+1,-1} + t \psi_{i-1,-1},
\label{eq:spin1-decouple-m1}
\end{multline}
\end{subequations}
%---------------------------------------------------------------------
Now we can choose $\epsilon_{i,1}=\epsilon_{i,0}=\epsilon_{i,-1}=\epsilon_{i}$, \ie, the values of the on-site 
potentials for the three different spin components at an $i$-th atomic site are assumed to be the same. We know that, 
for our model, the values of the local magnetic fields $h_{i}=h_{\mathcal{A}}$ and $h_{\mathcal{B}}$, are distributed 
following a Fibonacci pattern. We can choose exactly the same Fibonacci sequence for the values of the on-site potentials 
$\epsilon_{i}=\epsilon_{\mathcal{A}}$ and $\epsilon_{\mathcal{B}}$. We can then suitably control the values of 
$\epsilon_{\mathcal{A}}$ and $\epsilon_{\mathcal{B}}$ through some external gate voltages to have the 
appropriate correlation condition for the spin filtering. The possible correlation conditions for the spin-1 particles are 
$\epsilon_{i} = \Delta \pm h_{i},\ i \in \{\mathcal{A},\mathcal{B}\}$. 
%######################################################
\begin{figure}[ht]
\includegraphics[clip,width=0.49\columnwidth]{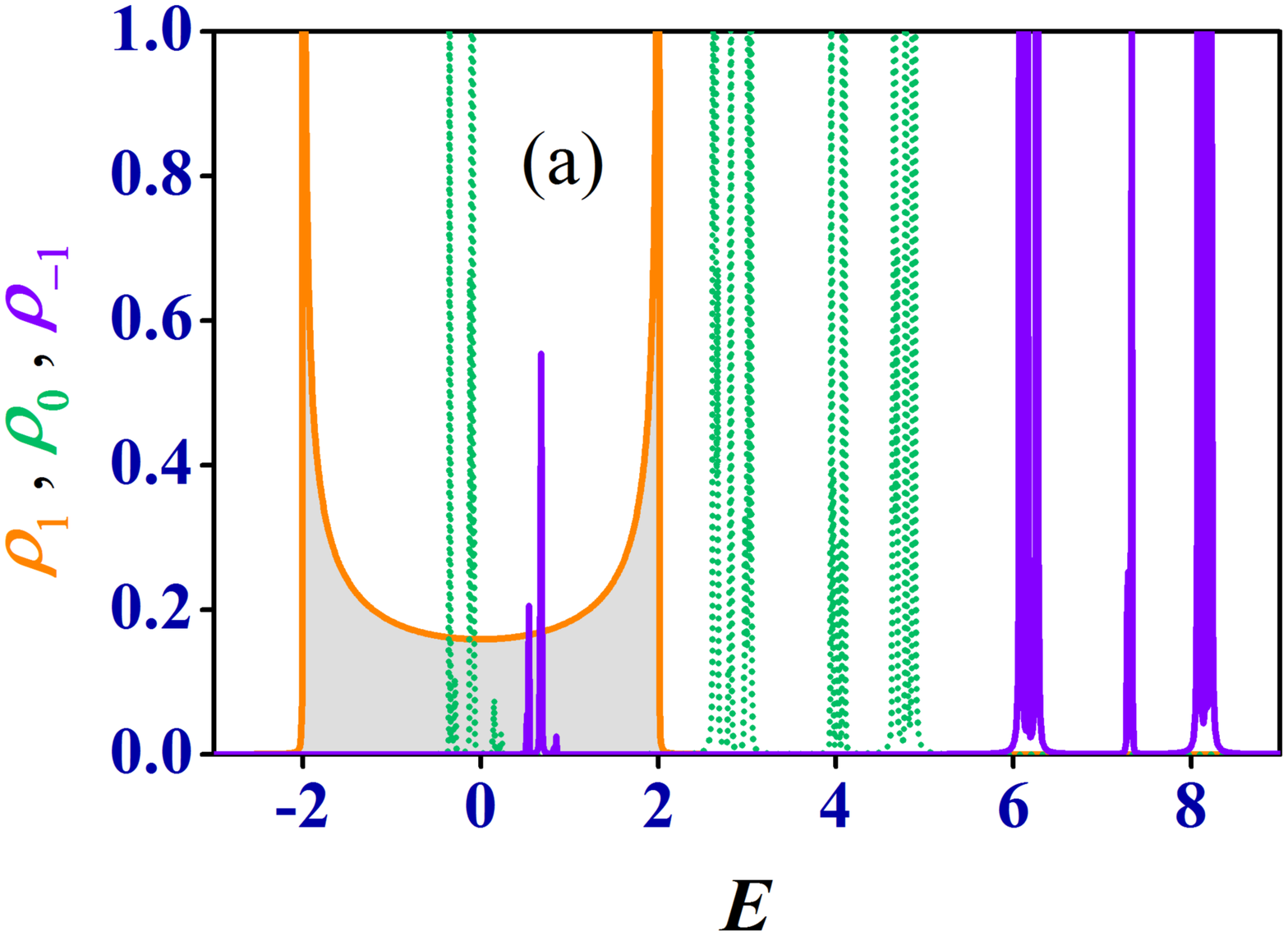}
\includegraphics[clip,width=0.49\columnwidth]{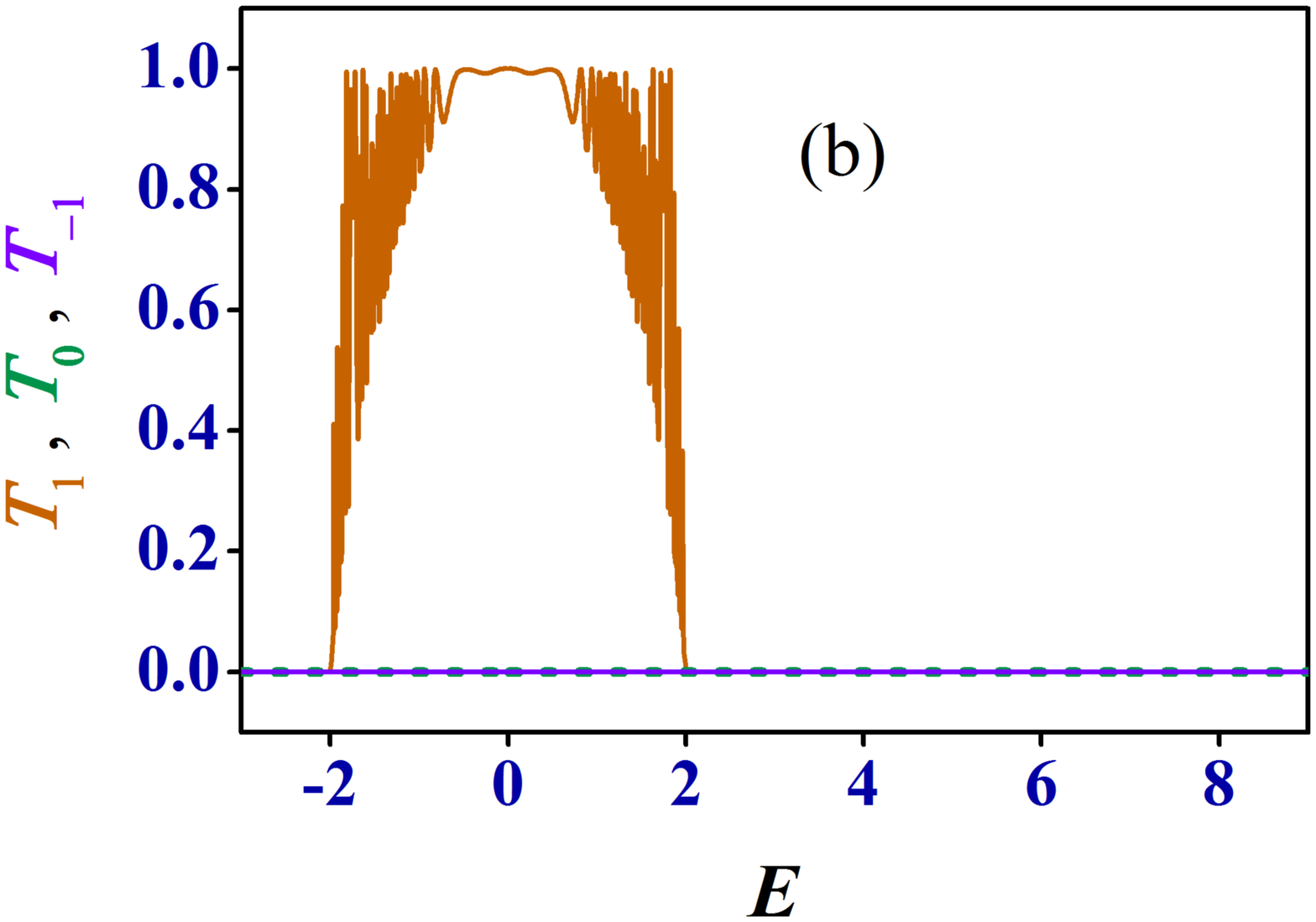}
\caption{(a) Plots for the LDOS for the spin-$1$, spin-$0$ and spin-$(-1)$ 
components with the correlation condition 
$\epsilon_{\mathcal{A}}=\Delta + h_{\mathcal{A}}$ 
and $\epsilon_{\mathcal{B}}=\Delta + h_{\mathcal{B}}$. 
We take $\Delta=0$, $h_{\mathcal{A}}=3.5$, and $h_{\mathcal{B}}=0.5$ 
measured in units of the hopping integral $t$. (b) The corresponding 
transmission probabilities $T_{1}$, $T_{0}$, and $T_{-1}$ evaluated 
for a $15$-th generation Fibonacci magnetic chain containing 
$610$ atomic sites. The lead parameters are 
$\epsilon_{S}=\epsilon_{D}=0$, $t_{S}=t_{D}=5$, and 
$t_{S,MC}=t_{MC,D}=4$ respectively. The plots with the dotted lines 
in both panels are for the spin-$0$ component.}
\label{fig:ldos-trans-spin1-1}
\end{figure}
%###################################################### 
By applying one of these two sets of conditions, 
we generate an absolutely continuous energy spectrum for one of the three spin channels while the remaining 
channels will have multifractal energy spectrum. Consequently, we will have one of the spin components passing 
through the system while the remaining ones will be completely blocked. In Fig.~\ref{fig:ldos-trans-spin1-1}, we 
exhibit one such situation as a prototype example, where we apply the correlation conditions 
$\epsilon_{\mathcal{A}} = \Delta + h_{\mathcal{A}}$ and $\epsilon_{\mathcal{B}} = \Delta + h_{\mathcal{B}}$. 
This makes the system completely transparent for the particles with spin-$1$ component, while completely impeding the 
particles with spin-$0$ and spin-$(-1)$ components. It is easily understood that, the other correlation condition will make 
the spin-$(-1)$ component transparent through the system while blocking the particles with spin-$0$ and spin-$1$ 
components. We do not show this result to save space. 
We note that, we cannot have spin filtering for the spin-$0$ component, as for the spin-$0$ channel, 
we do not have an `$h$ term' in the effective on-site potentials as reflected in Eq.~\eqref{eq:spin1-decouple-0}.
\subsection{Other higher-order spin systems}
It can be appreciated that, the mathematical framework we have used to compute the results for the previous 
two cases of a spin-half ($S=1/2$) and a spin-$1$ ($S=1$) system, can be easily extended for any general `spin-$S$' system. 
It is automatically understood that, as we go to the higher-order spin cases, we will have more numbers of spin components 
and eventually one of them will filter out through the system for the suitable correlations between the values of the 
local magnetic fields and the on-site potentials. Of course, we need to have different choices of correlations between 
$\epsilon_{i}$ and $h_{i}$ to achieve the spin filtering for different cases as we move up along the higher-order spin ladder. 
We note that, to have the filtering for a particle with certain spin component, the Fermi energy of the particle has to 
lie within a certain energy range where the absolutely continuum spectrum appears.  
One can easily work out that, 
for the spin case $S=3/2$, the set of correlation conditions will be 
$\epsilon_{i} = \Delta \pm h_{i}$ and $\epsilon_{i} = \Delta \pm (h_{i}/3)$; 
for the spin case $S=2$, the set of correlation conditions will be 
$\epsilon_{i} = \Delta \pm h_{i}$ and $\epsilon_{i} = \Delta \pm (h_{i}/2)$; 
and for the spin case $S=5/2$, the set of correlation conditions will be 
$\epsilon_{i} = \Delta \pm h_{i}$, $\epsilon_{i} = \Delta \pm (h_{i}/5)$, and 
$\epsilon_{i} = \Delta \pm (3h_{i}/5)$. 
Here $i \in \{\mathcal{A},\mathcal{B}\}$ for our Fibonacci quasiperiodic magnetic chain model. 
One can also calculate the conditions for the spin filtering for other higher-order spin particles following the 
same prescription. 

{\it Remark on the various disorder effects on the spin filtering phenomena.---} It is important to discuss the 
robustness of the spin filtering protocol with respect to various disorder effects in the system. One can easily 
understand that disorder or thermal fluctuations in the system will spoil the perfect alignment of the polar angles 
$\theta_{i} = 0$ $\forall$ $i$ of the magnetic moments. Hence, it is significant to understand what should be the cut-off 
limit in the random tilting of the $\theta_{i}$ angles before the spin filtering mechanism in the system breaks down. Upon 
performing a rigorous numerical investigation, it has been found that our results on the spin filtering effect are fairly 
robust for small random tilting of the angle $\theta_{i}$ below $\theta_{i} = \pm 10$ degrees. Beyond this critical value 
of random tilting of $\theta_{i}$, the states populating the energy spectrum consist of highly localized states, 
directing the system to act as a poorly conducting system. So it can be concluded that, for a weak disorder (or thermal 
broadening) corresponding to the above-mentioned critical value of random tilting of the angle $\theta_{i}$, the spin 
filtering protocol persists. 

Similarly, it is also worth addressing the effect of the mismatches between the on-site energies 
$\epsilon_{i}$ and the local magnetic fields $h_{i}$ at different atomic sites on the spin filtering effect. One can 
capture this effect by choosing a random $\Delta_{i}$. We have numerically found that for a very weak random disorder 
in $\Delta_{i}$, chosen randomly between the values $-0.2t$ and $0.2t$ ($t$ being the hopping amplitude), the spin 
filtering effect is preserved in the system. As we increase the strength of this disorder to higher values, strong 
localization effect starts to take over and the efficiency of the spin transport through the system is extensively 
reduced. So the protocol for the spin filtering in our system is robust against weak disorder in $\Delta_{i}$. 
It is also to be noted that, for larger $S$ the system could be more inclined towards thermal fluctuations. 
But in an actual experimental situation, such thermal fluctuations can be controlled by performing the experiment 
at a low temperature. However, we have to keep in mind that, from the point of view of practical applicability, 
one cannot go down too low in the temperature scale. Hence, for a real-life application purpose one has 
to judiciously compromise between the thermal fluctuations and the temperature scale. 
%%%%%%%%%%%%%%%%%%%%%%%%%%%%%%%%%%%%%%%%%%%%%%%%%%%%%%%%%%%%%%%%
\section{Conclusion}
\label{conclu}
In this paper, we have studied the spin-dependent transport for particles with arbitrary higher-order spin 
states in a tight-binding quasiperiodic magnetic chain model. A mathematical analogy between a 
multi-strand ladder network for spinless particles and a multicomponent spin-$S$ system mimicking a 
$(2S+1)$ strand ladder network has been exploited to analyze the problem and extract the useful results 
for the spin filtering for different spin components. We show that, by incorporating a suitable correlation 
between the magnitude of the of local magnetic fields and the on-site potentials of the magnetic atomic 
sites, one can render an absolutely continuous energy spectrum for one of the desired spin 
channels (components) with a highly transmitting window for the entire allowed energy range, while 
the other spin channels exhibit a multifractal energy spectrum with zero transmission probabilities. We 
show and explain the results in detail for two prototype examples of spin-half ($S=1/2$) and spin-$1$ 
($S=1$) cases. We also give the outline for the other higher order spin cases and justify that, the 
essential mathematical exercise employed by us for our problem is a general one that holds true for 
any arbitrary higher-order spin-$S$ particles, where $S$ is an integer or half-integer. 

Some of the recent interesting experimental studies~\cite{wiesendanger-nn10, wiesendanger-np12} 
show that, it is possible to manipulate the spin direction of individual magnetic atoms to form nanomagnets 
with arrays of a few exchange-coupled atomic magnetic moments, exhibiting a rich variety of magnetic 
properties and can be explored as the constituents of nanospintronics technologies. This indicates 
that, our theoretical proposal of a quasiperiodic magnetic chain with an array of atomic 
magnetic moments sequenced in a Fibonacci pattern is not far from reality and might be realized in real-life 
experiments. Our results can be useful to realize novel magnetic quantum information storage 
devices and spin-based logic operators~\cite{wiesendanger-sci11}, relying on the operation of higher-order 
spin states. One can carry forward our analysis and results of this work for systems with other 
quasiperiodic sequences like Thue-Morse, period-doubling, copper mean etc. 
Finally, we believe that, our theoretical study of the realization of spin filters by using a simple 
tight-binding quasiperiodic magnetic chain system might open up an interesting futuristic prospect of 
realizing spin filters using quasicrystalline materials. 
%%%%%%%%%%%%%%%%%%%%%%%%%%%%%%%%%%%%%%%%%%%%%%%%%%%%%%%%%%%%%%%%           
\begin{acknowledgments}
The author gratefully acknowledges the funding and the facilities provided by MPIPKS through a postdoctoral 
scholarship. The author would like to thank A. Mukherjee and T. Nag for some useful suggestions on the 
manuscript. The author would also like to express his appreciation for the valuable suggestions and 
constructive criticisms of the anonymous referees in preparing the revised manuscript. 
\end{acknowledgments}
%%%%%%%%%%%%%%%%%%%%%%%%%%%%%%%%%%%%%%%%%%%%%%%%%%%%%%%%%%%%%%%%


\begin{thebibliography}{99}

\bibitem{prinz-science98} G. Prinz, \href{http://doi.org/10.1126/science.282.5394.1660}
{Science \textbf{282}, 1660 (1998)}.

\bibitem{wolf-sci01} S. A. Wolf, D. D. Awschalom, R. A. Buhrman, J. M. Daughton, 
S. von Moln\'{a}r, M. L. Roukes, A. Y. Chtchelkanova, and D. M. Treger, 
\href{http://doi.org/10.1126/science.1065389}{Science \textbf{294}, 1488 (2001)}.

\bibitem{sahoo-np05} S. Sahoo, T. Kontos, J. Furer, C. Hoffmann, M. Gr\"{a}ber, 
A. Cottet, and C. Sch\"{o}nenberger, 
\href{https://doi.org/10.1038/nphys149}{Nat. Phys. \textbf{1}, 99 (2005)}.

\bibitem{datta-prl02} T. Koga, J. Nitta, H. Takayanagi, and S. Datta, 
\href{https://doi.org/10.1103/PhysRevLett.88.126601}{Phys. Rev. Lett. \textbf{88}, 126601 (2002)}.

\bibitem{umansky-sci03} J. A. Folk, R. M. Potok, C. M. Marcus and V. Umansky, 
\href{https://doi.org/10.1126/science.1078419}{Science \textbf{299}, 679 (2003)}.

\bibitem{west-prl04} L. P. Rokhinson, V. Larkina, Y. B. Lyanda-Geller, L. N. Pfeiffer, and K. W. West, 
\href{https://doi.org/10.1103/PhysRevLett.93.146601}{Phys. Rev. Lett. \textbf{93}, 146601 (2004)}.

\bibitem{umansky-prl03} S. K. Watson, R. M. Potok, C. M. Marcus, and V. Umansky, 
\href{https://doi.org/10.1103/PhysRevLett.91.258301}{Phys. Rev. Lett. \textbf{91}, 258301 (2003)}.

\bibitem{loss-prl00} P. Recher, E. V. Sukhorukov, and D. Loss, 
\href{https://doi.org/10.1103/PhysRevLett.85.1962}{Phys. Rev. Lett. \textbf{85}, 1962 (2000)}.

\bibitem{kim-apl01} A. A. Kiseleva and K. W. Kim, 
\href{https://doi.org/10.1063/1.1347023}{Appl. Phys. Lett. \textbf{78}, 775 (2001)}. 

\bibitem{leclair-apl02} P. LeClair, J. K. Ha, H. J. M. Swagten, J. T. Kohlhepp, 
C. H. van de Vin, and W. J. M. de Jonge, 
\href{https://doi.org/10.1063/1.1436284}{Appl. Phys. Lett. \textbf{80}, 625 (2002)}.

\bibitem{brandbyge-prl07} M. Koleini, M. Paulsson, and M. Brandbyge, 
\href{https://doi.org/10.1103/PhysRevLett.98.197202}{Phys. Rev. Lett. \textbf{98}, 197202 (2007)}.

\bibitem{feng-apl10} M. G. Zeng, L. Shen, Y. Q. Cai, Z. D. Sha, and Y. P. Feng, 
\href{https://doi.org/10.1063/1.3299264}{Appl. Phys. Lett. \textbf{96}, 042104 (2010)}.

\bibitem{chuang-nn15} P. Chuang, S.-C. Ho, L. W. Smith, F. Sfigakis, M. Pepper, C.-H. Chen, J.-C. Fan, 
J. P. Griffiths, I. Farrer, H. E. Beere, G. A. C. Jones, D. A. Ritchie, and T.-M. Chen, 
\href{https://doi.org/10.1038/nnano.2014.296}{Nat. Nanotechnol. \textbf{10}, 35 (2015)}.

\bibitem{yan-nc16} W. Yan, O. Txoperena, R. Llopis, H. Dery, L. E. Hueso, and F. Casanova, 
\href{https://doi.org/10.1038/ncomms13372}{Nat. Commun. \textbf{7}, 13372 (2016)}.

\bibitem{datta-apl90} S. Datta and B. Das, 
\href{https://doi.org/10.1063/1.102730}{Appl. Phys. Lett. \textbf{56}, 665 (1990)}.

\bibitem{nitta-apl99} J. Nitta, F. E. Meijer, and H. Takayanagi, 
\href{https://doi.org/10.1063/1.124485}{Appl. Phys. Lett. \textbf{75}, 695 (1999)}

\bibitem{loss-book02} \textit{Semiconductor Spintronics and Quantum Computation}, 
edited by D. Awschalom, D. Loss, and N. Samarth  (Springer, New York, 2002).

\bibitem{nagasawa-prl12} F. Nagasawa, J. Takagi, Y. Kunihashi, M. Kohda, and J. Nitta, 
\href{https://doi.org/10.1103/PhysRevLett.108.086801}{Phys. Rev. Lett. \textbf{108}, 086801 (2012)}.

\bibitem{richter-prb04} D. Frustaglia and K. Richter, 
\href{https://doi.org/10.1103/PhysRevB.69.235310}{Phys. Rev. B \textbf{69}, 235310 (2004)}.

\bibitem{fiederling-nat99} R. Fiederling, M. Keim, G. Reuscher, W. Ossau, G. Schmidt, A. Waag, and L.W. Molenkamp, 
\href{https://doi.org/10.1038/45502}{Nature (London) \textbf{402}, 787 (1999)}.

\bibitem{ohno-nat99} Y. Ohno, D. K. Young, B. Beschoten, F. Matsukura, H. Ohno, and D. D. Awschalom, 
\href{https://doi.org/10.1038/45509}{Nature (London) \textbf{402}, 790 (1999)}.

\bibitem{bauer-prb02} Y. Tserkovnyak, A. Brataas, and G. E. W. Bauer, 
\href{https://doi.org/10.1103/PhysRevB.66.224403}{Phys. Rev. B \textbf{66}, 224403 (2002)}

\bibitem{hammar-prl9902} P. R. Hammar, B. R. Bennett, M. J. Yang, and M. Johnson, 
\href{https://doi.org/10.1103/PhysRevLett.83.203}{Phys. Rev. Lett. \textbf{83}, 203 (1999)}; 
P. R. Hammar and M. Johnson, 
\href{https://doi.org/10.1103/PhysRevLett.88.066806}{Phys. Rev. Lett. \textbf{88}, 066806 (2002)}.

\bibitem{popp-nanotech03} M. Popp, D. Frustaglia, and K. Richter, 
\href{https://doi.org/10.1088/0957-4484/14/2/347}{Nanotechnol. \textbf{14}, 347 (2003)}.

\bibitem{bercioux-prl04} D. Bercioux, M. Governale, V. Cataudella, and V. M. Ramaglia, 
\href{https://doi.org/10.1103/PhysRevLett.93.056802}{Phys. Rev. Lett. \textbf{93}, 056802 (2004)}.
 
\bibitem{aharony-prb08} A. Aharony, O. Entin-Wohlman, Y. Tokura, and S. Katsumoto, 
\href{https://doi.org/10.1103/PhysRevB.78.125328}{Phys. Rev. B \textbf{78}, 125328 (2008)}.

\bibitem{pal-scirep16} B. Pal and P. Dutta, 
\href{https://doi.org/10.1038/srep32543}{Sci. Rep. \textbf{6}, 32543 (2016)}.

\bibitem{pan-prb15} T.-R. Pan, A.-M. Guo, and Q.-F. Sun, 
\href{https://doi.org/10.1103/PhysRevB.92.115418}{Phys. Rev. B \textbf{92}, 115418 (2015)}.

\bibitem{guo-prl12} A.-M. Guo and Q.-F. Sun, 
\href{https://doi.org/}{Phys. Rev. Lett. \textbf{108}, 218102 (2012)}.

\bibitem{gohler-sci11} B. G\"{o}hler, V. Hamelbeck, T. Z. Markus, M. Kettner, 
G. F. Hanne, Z. Vager, R. Naaman, and H. Zacharias, 
\href{https://doi.org/10.1126/science.1199339}{Science \textbf{331}, 894 (2011)}.

\bibitem{pal-jpcm16} B. Pal, R. A. R\"{o}mer, and A. Chakrabarti, 
\href{https://doi.org/10.1088/0953-8984/28/33/335301}{J. Phys.: Condens. Matter \textbf{28}, 335301 (2016)}.

\bibitem{ho-prl99} T.-L. Ho and S. Yip, 
\href{https://doi.org/10.1103/PhysRevLett.82.247}{Phys. Rev. Lett. \textbf{82}, 247 (1999)}. 

\bibitem{azaria-prl05} P. Lecheminant, E. Boulat, and P. Azaria, 
\href{https://doi.org/10.1103/PhysRevLett.95.240402}{Phys. Rev. Lett. \textbf{95}, 240402 (2005)}.

\bibitem{pfau-np06} M. Fattori, T. Koch, S. Goetz, A. Griesmaier, S. Hensler, J. Stuhler, and T. Pfau, 
\href{https://doi.org/10.1038/nphys443}{Nat. Phys. \textbf{2}, 765 (2006)}.

\bibitem{rey-np10} A. V. Gorshkov, M. Hermele, V. Gurarie, C. Xu, P. S. Julienne, 
J. Ye, P. Zoller, E. Demler, M. D. Lukin, and  A. M. Rey, 
\href{https://doi.org/10.1038/nphys1535}{Nat. Phys. \textbf{6}, 289 (2010)}.

\bibitem{takahasi-np12} S. Taie, R. Yamazaki, S. Sugawa, and Y. Takahashi, 
\href{https://doi.org/10.1038/nphys2430}{Nat. Phys. \textbf{8}, 825 (2012)}.

\bibitem{pagano-np14} G. Pagano, M. Mancini, G. Cappellini, P. Lombardi, F. Sch\"{a}fer, H. Hu, 
X.-J. Liu, J. Catani, C. Sias, M. Inguscio, and L. Fallani, 
\href{https://doi.org/10.1038/nphys2878}{Nat. Phys. \textbf{10}, 198 (2014)}. 

\bibitem{kohmoto-prl83} M. Kohmoto, L. P. Kadanoff, and C. Tang, 
\href{https://doi.org/10.1103/PhysRevLett.50.1870}{Phys. Rev. Lett. \textbf{50}, 1870 (1983)}.

\bibitem{kohmoto-prb86} C. Tang and M. Kohmoto, 
\href{https://doi.org/10.1103/PhysRevB.34.2041}{Phys. Rev. B \textbf{34}, 2041(R) (1986)}. 

\bibitem{kittel-book} C. Kittel, \textit{Introduction to Solid State Physics}, 8th ed.,
(Wiley, New York, 2005).

\bibitem{anderson-pr58} P. W. Anderson, 
\href{https://doi.org/10.1103/PhysRev.109.1492}{Phys. Rev. \textbf{109}, 1492 (1958)}.

\bibitem{mukherjee-prb18} A. Mukherjee, A. Chakrabarti, and R. A. R\"{o}mer, 
\href{https://doi.org/10.1103/PhysRevB.98.075415}{Phys. Rev. B \textbf{98}, 075415 (2018)}.

\bibitem{sil-prb08} S. Sil, S. K. Maiti, and A. Chakrabarti, 
\href{https://doi.org/10.1103/PhysRevB.78.113103}{Phys. Rev. B \textbf{78}, 113103 (2008)}.

\bibitem{pal-physE14} B. Pal and A. Chakrabarti, 
\href{https://doi.org/10.1016/j.physe.2014.02.022}{Physica E \textbf{60}, 188 (2014)}.

\bibitem{pal-pla14} B. Pal and A. Chakrabarti, 
\href{https://doi.org/10.1016/j.physleta.2014.07.034}{Phys. Lett. A \textbf{378}, 2782 (2014)}.

\bibitem{wiesendanger-nn10} D. Serrate, P. Ferriani, Y. Yoshida, S.-W. Hla, M. Menzel, 
K. von Bergmann, S. Heinze, A. Kubetzka, and R. Wiesendanger,
\href{https://doi.org/10.1038/nnano.2010.64}{Nat. Nanotechnol. \textbf{5}, 350 (2010)}.

\bibitem{wiesendanger-np12} A. A. Khajetoorians, J. Wiebe, B. Chilian, S. Lounis, S. Bl\"{u}gel, 
and R. Wiesendanger, 
\href{https://doi.org/10.1038/nphys2299}{Nat. Phys. \textbf{8}, 497 (2012)}.

\bibitem{wiesendanger-sci11} A. A. Khajetoorians, J. Wiebe, B. Chilian, and R. Wiesendanger, 
\href{https://doi.org/10.1126/science.1201725}{Science \textbf{332}, 1162 (2011)}.

%\bibitem{} 

%\bibitem{} 

%\bibitem{} 

\end{thebibliography}
\end{document}